\documentclass{article}

\usepackage{amsmath}  % load first
\usepackage[english]{babel}
\usepackage[utf8]{inputenc}
\usepackage{amssymb}
\usepackage{graphicx}
\usepackage{authblk}
\usepackage{caption}
\usepackage{multirow}
\usepackage{amscd}
\usepackage{geometry}
\usepackage{a4wide}
\usepackage[usenames]{color}
\usepackage{etaremune} 
\usepackage[hidelinks]{hyperref}
\usepackage{amsfonts}
\usepackage{ifthen}
\usepackage{color}
\usepackage{amscd}
\usepackage{latexsym}
\usepackage{xfrac} % per sfrac

\usepackage{tikz}
\usepackage{tikz-3dplot}

\usepackage{comment}

\newcommand{\bd}{\begin{definition}}
	\newcommand{\ed}{\end{definition}}
\newcommand{\bt}{\begin{theorem}}
	\newcommand{\et}{\end{theorem}}
\newcommand{\bi}{\begin{itemize}}
	\newcommand{\ei}{\end{itemize}}
\newcommand{\ben}{\begin{enumerate}}
	\newcommand{\een}{\end{enumerate}}
\newcommand{\beq}{\begin{equation}}
\newcommand{\eeq}{\end{equation}}

\newcommand{\R}{\mbox{$ \mathbb{R}  $}}
\newcommand{\C}{\mbox{$ \mathbb{C} $}}

\newcommand{\pmat}[1]{\begin{pmatrix} #1 \end{pmatrix}}

\newcommand{\ket}[1]{\vert {#1} \rangle}
\newcommand{\Tr}{\text{Tr}}
\newcommand*\bell{\ensuremath{\boldsymbol\ell}}

\newtheorem{definition}{Def.}[section]
\newtheorem{theorem}{Theorem}[section]

\def\keywords{\vspace{.5em}
	{\noindent \textbf{Keywords}:\,\relax%
}}

\begin{document}

\title{An Operational Quantum Information Framework for Experimental Studies on Color Perception}

\author[1]{Roberto Leporini\thanks{roberto.leporini@unibg.it}}
\author[2]{Edoardo Provenzi\thanks{edoardo.provenzi@math.u-bordeaux.fr}}
\author[3]{Michel Berthier\thanks{michel.berthier@univ-lr.fr}}
%\author[2]{Author D\thanks{D.D@university.edu}}
%\author[2]{Author E\thanks{E.E@university.edu}}
\affil[1]{Department of Economics, University of Bergamo, via dei Caniana 2, 24127, Bergamo, Italy}
\affil[2]{Université de Bordeaux, CNRS, Bordeaux INP, IMB, UMR 5251\\ F-33400, 351 Cours de la Libération, Talence, France}
\affil[3]{Laboratoire MIA, Batiment Pascal, Pôle Sciences et Technologie, Université de La Rochelle, 23, Avenue A. Einstein, BP 33060, 17031 La Rochelle cedex, France}

\renewcommand\Authands{ and }
\date{\today}

\maketitle

%\begin{history}
%\received{(Day Month Year)}
%\revised{(Day Month Year)}
%\end{history}

\begin{abstract}
	Starting from the foundational axiomatization of the perceptual color space initiated by Schrödinger in 1920 and eventually refined by Resnikoff in 1974, Berthier, Provenzi and their collaborators have recently proposed a reformulation of perceptual color attributes within the framework of quantum information. Their work is 
	based on the Jordan algebra formalism of quantum theories and, more specifically, on a quantum system described by a spin factor over the field of real numbers. This theoretical framework is not that of ordinary quantum mechanics, mainly because it requires dealing with rebits, whereas the latter uses qubits. The aim of this paper is to show that this difference in no way hinders the implementation of experimental protocols for testing the validity of the predictions of the color perception model. In particular, we show how to compute the quantum information based perceptual attributes of perceived colors in terms of qubit density matrices. 
\end{abstract}


	\section{Introduction}

In the paper \cite{Berthier:22SIAM}, a mathematically rigorous vocabulary for color attributes has been built based on a prior analysis of color perception. This analysis, presented in a series of papers, provided theoretical evidence supporting the development of a quantum-relativistic theory of color perception free from internal contradictions. For more details, see e.g.  \cite{Berthier:2020, Berthier:2021JofImaging, Berthier:21JMP, BerthierProvenzi:2022PRS}.

A distinctive feature of this theory is that, unlike the majority of quantum models, the state space is defined over the real field rather than the complex one. If fact, the quantum system of this theory is a \textit{rebit}, the real analogue of a qubit. In this model, the two states correspond to the degrees of chromatic opposition as described in Hering's theory of color vision \cite{Hering:1878}. To the best of our knowledge, this color perception model is the first factual example of a real quantum theory, which had previously been explored only theoretically.

It is a challenging problem to validate this new theory experimentally, for example by testing some of its predictions. In tests designed to characterize a color perceived by a human observer, photons of visible light are absorbed by the retinal photoreceptors, changing their electrical potential and triggering a very complicated chain of events that culminate in the brain's creation of a color sensation. The perceptual attributes associated with this sensation are defined in the proposed theory from the generalized state of the rebit given by the result of a Lüders measurement operation.
Therefore, in order to perform experiments, we must take into account the inherent fact that photons are described in quantum mechanics using qubits, while color sensations are described in the quantum model of color perception using rebits.

Replacing the field of complex numbers of ordinary quantum mechanics by the field of real numbers is not a trivial matter. For instance, it is underlined in \cite{Wootters:1990} that in real-vector-space quantum mechanics, one cannot in general determine the density matrix of a composite system using only local measurements whereas in complex-vector-space quantum mechanics any sets of measurements which are just sufficient for determining the states of the subsystems are, when performed jointly, just sufficient for determining the state of the complete system.

Another example is highlighted in \cite{Alde:2023}, where it is shown that Choi's theorem cannot be applied in the real-vector-space framework to obtain the classification of rebit channels and another strategy must be used.

Nevertheless, if we limit the scope of the experiments to be performed, it is possible to express some of the colorimetric concepts defined in the rebit framework by means of density matrices associated with qubit states. It is precisely the aim of this paper to elaborate on this last point.

The paper is structured as follows: in section \ref{sec:recap}, we recall the basic facts about the quantum model of color perception together with the quantum information-based definitions of the perceptual attributes that characterize perceived colors.

In section \ref{sec:recast}, we explain how to compute these perceptual attributes using qubit states and effects, rather than using rebit generalized states and effects. In doing so, we propose an operational framework for conducting color perception experiments. 

Finally, in the conclusions we discuss some perspectives for designing such experiments that we deem able to test the quantum color perception model.

\section{Hering's rebit of color perception}\label{sec:recap}

The thorough historical, physiological, and psychophysical motivations, as well as the main stages of the mathematical reasoning that led to the proposal of the model of color perception described below, have been detailed in previous works, notably in \cite{Berthier:22SIAM} and \cite{BerthierProvenzi:2022PRS}.

Before delving into the technical description of the quantum information-based model of color perception, we briefly recall the core motivation behind this novel framework.

During the late 19th century, prominent scientists such as Riemann, Maxwell, Grassmann, and von Helmholtz began to recognize that the set of perceptual colors, which we denote by $\mathcal{C}$, should not be understood merely as a collection of subjective sensations, but as a space with a well-defined and nontrivial mathematical structure.

It was Schrödinger, in his foundational work \cite{Schroedinger:20}, who first formalized this intuition by establishing a set of axioms showing that $\mathcal{C}$ forms a 3-dimensional regular convex cone. This geometric insight laid the groundwork for subsequent axiomatic approaches to color perception.

Decades later, Resnikoff refined Schrödinger’s analysis in \cite{Resnikoff:74}, proving that $\mathcal{C}$ is not only a convex cone but also a homogeneous space. This result implies that $\mathcal{C}$ must be isomorphic to one of two canonical spaces: either $[0,+\infty)^3$ or $[0,+\infty)\times \mathbf{H}$, where $\mathbf{H}$ is a two-dimensional hyperbolic space. 

For reasons that will soon become apparent, it is important to emphasize that $\mathcal{C}$ represents an idealized perceptual space, meaning it includes all virtually possible color sensations in isolation, independently of the physiological thresholds imposed by human vision at very low or very large light intensities.

The Cartesian space $[0,+\infty)^3$ is the underlying geometric structure of standard colorimetric models such as those promoted by the CIE (Commission Internationale de l'Éclairage). However, these models are not well suited to capture the intrinsically perceptual aspects of color experience. In contrast, the space $[0,+\infty)\times \mathbf{H}$ exhibits significantly richer algebraic, geometric, and perceptual properties. Therefore, in the rest of this work, we focus exclusively on this second option.

Notably, the space $[0,+\infty)\times \mathbf{H}$ is isomorphic to the cone of $2\times2$ real symmetric positive-semidefinite matrices, which in turn is isomorphic to $\overline{\mathcal{L}^+}$, the closure of the future lightcone in the 3-dimensional Minkowski space.

In his concluding remarks, Resnikoff observed that these spaces also coincide with the \textit{domains of positivity} of the only two 3-dimensional, non-associative, formally real Jordan algebras\footnote{see, e.g., \cite{Baez:12} for more information about Jordan algebras.}: $\mathcal{H}(2,\mathbb{R})$, the Jordan algebra of real symmetric $2\times2$ matrices equipped with the Jordan product 
\beq
A \circ B := \frac{1}{2}(AB + BA),
\eeq 
and $\mathbb{R} \oplus \mathbb{R}^2$, known as the \textit{spin factor}, endowed with the Jordan product
\beq
(\alpha,\mathbf{v}) \circ (\beta,\mathbf{w}) := \left(\alpha\beta + \mathbf{v} \cdot \mathbf{w},\; \alpha \mathbf{w} + \beta \mathbf{v}\right),\eeq
for $\alpha, \beta \in \mathbb{R}$ and $\mathbf{v}, \mathbf{w} \in \mathbb{R}^2$, where $\mathbf{v} \cdot \mathbf{w}$ denotes the standard Euclidean inner product. These two Jordan algebras are isomorphic via the map:
\beq\label{eq:isoH2}
\begin{array}{cccl}
	\chi: & \mathcal H(2,\mathbb R)  & \stackrel{\sim}{\longrightarrow} & \mathbb R\oplus \mathbb R^2 \\
	& 	A=\begin{pmatrix}
		\alpha + v_1 & v_2 \\
		v_2 & \alpha - v_1
	\end{pmatrix}  & \longmapsto & \chi(A)=  \begin{pmatrix} \alpha\\\textbf{v} \end{pmatrix}, \qquad \textbf{v}=\pmat{v_1 \\ v_2}.
\end{array}
\eeq

The quantum information-theoretic model we introduce in the next section is based on the assumption that these naturally emerging Jordan algebras encode the structure of the quantum observables relevant to color perception. This identification provides both a geometrically grounded and algebraically coherent foundation for our approach.

\subsection{Perceived colors and measurements in the quantum  framework}

%	The relevant algebraic structure of this model is the spin factor  $\R\oplus \R^2$ whose Jordan product is given by $(\alpha,\textbf{v})\circ (\beta,\textbf{w})=(\alpha\beta + \textbf{v} \cdot \textbf{w}, \alpha \textbf{w} + \beta \textbf{v})$, with $\alpha,\beta \in \R$ and $\textbf{v},\textbf{w}\in \R^2$, and where $\textbf{v}\cdot \textbf{w}$ denotes the Euclidean inner product of $\R^2$. It is a non-associative commutative formally real Jordan algebra of rank 2 and of real dimension 3 which, as a Jordan algebra, is 

The trace of an element $(\alpha,\textbf{v})$ of $\R\oplus \R^2$ is equal to $2\alpha$, so the unit trace elements of the spin factor are in one-to-one correspondence via the isomorphism $\chi$ with the density matrices of the rebit, i.e. the unit trace matrices belonging to the domain of positivity $\overline{\cal H^+}(2,\R)$ of the algebra $\mathcal H(2,\R)$. We denote
\beq\label{eq:dmat}
\mathcal S = \left\{ \rho(s_1,s_2) \equiv \frac{1}{2}\begin{pmatrix}
	1+s_1 & s_2 \\ s_2 & 1-s_1
\end{pmatrix}, \;  s_1,s_2\in \R, \; s_1^2+s_2^2\le 1 \right\} \cong \mathcal D,
\eeq 
the state space of the rebit, where $\cal D$ is the unit disk in $\R^2$, referred to as the Bloch disk.

Vectors of the Bloch disk represent chromatic states, and Hering's color opponency involving the two pairs of unique hues is encoded by means of the two Pauli matrices with real entries 
\begin{equation}
	\sigma_1=\begin{pmatrix}
		1 & 0 \\ 0 & -1
	\end{pmatrix}, \; \sigma_2=\begin{pmatrix}
		0 & 1 \\ 1 & 0
	\end{pmatrix}.
\end{equation}
In the sequel, we will denote with $\sigma_0$ the $2\times 2$ identity matrix $I_2$.

As in the case of qubits, we can decompose any density matrix of the rebit using the basis $(\sigma_0,\sigma_1,\sigma_2)$:
\begin{equation}
	\rho(s_1,s_2)={\rho_0}+\frac{s_1} 2\sigma_1+\frac{s_2} 2\sigma_2\equiv \rho_0 + \frac 1 2 \textbf{v}_\rho \cdot \vec \sigma,
\end{equation}
where $\rho_0:=\sigma_0/2$ and the components of the Bloch vector \beq
\textbf{v}_\rho=(s_1,s_2)=(\Tr(\rho \sigma_1),\Tr(\rho \sigma_2))\equiv( \langle \sigma_1\rangle_\rho,\langle\sigma_2\rangle_\rho)
\eeq 
are the $\rho$-expectation values of the real Pauli matrices.
	Using polar coordinates, i.e. $(s_1,s_2)=(r\cos \vartheta,r\sin \vartheta)$, $r\in [0,1]$, $\vartheta\in [0,2\pi)$, we can write
	\begin{equation}\label{eq:densmat}
		\rho(r,\vartheta)={\rho_0}+\frac{\langle \sigma_1\rangle_\rho}{2}\left[\rho(1,0)-\rho(1,\pi)\right]+\frac{\langle \sigma_2\rangle_\rho}{2}\left[\rho\left(1,{\pi/ 2}\right)-\rho\left(1,{3\pi/ 2}\right)\right].
	\end{equation}
	For all $\vartheta$, $\rho(1,\vartheta)$ is a rank-1 projector, i.e. a pure state, and $\rho(1,\vartheta_1)$,  $\rho(1,\vartheta_2)$ project on orthogonal directions precisely when $\vartheta_1$ and $\vartheta_2$ correspond to antipodal points on the unit circle. Since orthogonality in quantum theories represents incompatible states, eq. \eqref{eq:densmat} codifies \textit{a generic chromatic state} as the \textit{superposition of two chromatic opponencies between incompatible states}, red-green and yellow-blue in Hering's theory, see \cite{Hering:1878}, weighted by the expectation values of the real density matrices, \textit{plus an offset state} represented by $\rho_0$. 
	
	The density matrix $\rho_0$, associated with the center of the Bloch disk $\cal D$, is the maximally mixed state, characterized by the fact of having maximal von Neumann entropy, defined as $S(\rho)=-\text{Tr}(\rho\log_2 \rho)$, and not carrying any chromatic information. Therefore, the density matrix $\rho_0$ represents the \textit{achromatic state}, and eq. \eqref{eq:densmat} is exactly the quantum description of the chromatic information that can be gathered from an isolated color stimulus in Hering's theory, see e.g. \cite{Hubel:95}.
	
	The new paradigm at the heart of the model of color perception is based on the interpretation of perceived colors as the results of measurements performed by observers from chromatic states. Perceived colors are no longer simple triplets of coordinates in color spaces derived from XYZ space, but are produced by the duality between states and effects.
	
	The domain of positivity of the spin factor $\mathbb R\oplus\mathbb R^2$ is the closure $\overline{\mathcal L^+}$ of the future lightcone in the 3-dimensional Minkowski space. Constraints such as the visibility threshold and the glare limit impose to restrict the set of positive observables to a convex solid of finite volume of $\overline{\mathcal L^+}$, called color solid in colorimetry.
	%As previously remarked, the observables belonging to the infinite cone $\overline{\mathcal L^+}$ and the chromatic states contained in the Bloch disk $\cal D$ represent an ideal situation in which the real limitations of human ability to perceive colors are not taken into account. However, factors like the visibility threshold and the glare limit imply that the space of perceived colors is a finite-volume convex subset, known as the color solid, contained in $\overline{\mathcal L^+}$.
	%Modern quantum information theory offers a surprisingly reach and well-suited set of tools to handle this and other issues properly. The main observation from which all the following results of this section descend is that \textit{a perceived color is the result of a (quantum) measurement} that a human observer performs by looking at a color stimulus. 
	%
	%As it is well-known, in the standard interpretation of quantum phenomena, the measurement process is intrinsically probabilistic. The modern mathematical tool that encodes this feature is called \textit{effect}, see e.g.  \cite{Kraus:83,Busch:97,Heinosaari:2011} for an overview on this fundamental object. 
	
\begin{figure}[htbp]
		\centering
		\minipage{0.32\textwidth}
		\includegraphics[width=\linewidth, scale = 0.5]{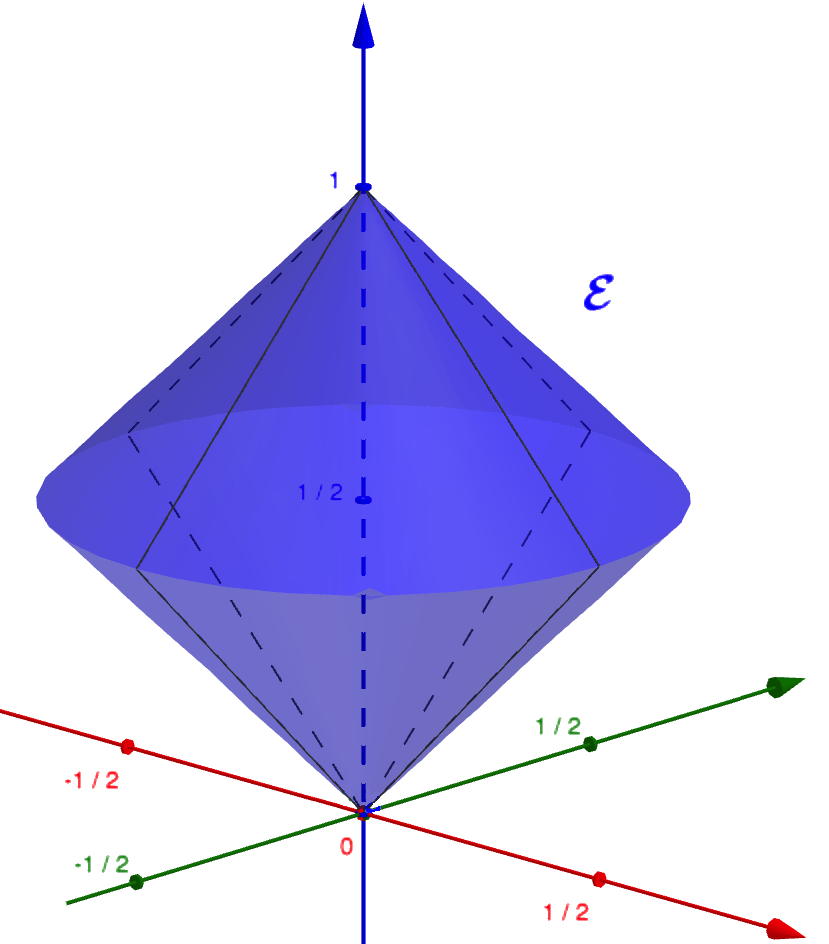}
		\endminipage
		\caption{The double cone representing the effect space can be interpreted as the color solid of actually perceived colors inside the infinite cone $\overline{\mathcal L^+}$.}
		\label{fig:E}
	\end{figure}

	Effects encode the probabilistic nature of quantum measurements and permit to construct in a natural way a color solid. To see how, we recall that an effect is defined to be an element $\eta_{\bf e}$ of $ \overline{\mathcal H^+}(2,\R)$ bounded between the null and the identity matrix $\sigma_0$ (with respect to the Loewner ordering of positive semi-definite matrices, i.e. $B\le A \iff A-B\in  \overline{\mathcal H^+}(2,\R)$). The matrix $\eta_{\bf e}$ can be written explicitly as follows
	\begin{equation}\label{eq:mateff}
		\eta_{\bf e}=\left(\begin{array}{cc}e_0+e_1 & e_2 \\e_2 & e_0-e_1\end{array}\right),
	\end{equation}
	with $e_0,e_1,e_2\in \R$ belonging to the following effect space:
	\begin{equation}\label{eq:effect1}
		\mathcal E=\left\{(e_0,e_1,e_2)\in \R^3, \; e_0 \in [0,1], \;  e_1^2+e_2^2\le \min \left\{(1-e_0)^2,e_0^2\right\} \right\}.
	\end{equation}
	The effect vector in the Bloch disk associated to $\eta_\textbf{e}$ is given by
	\beq\label{eq:veceff}
	{\bf v}_{\bf e}:=\left(\frac{e_1}{e_0},\frac{e_2}{e_0}\right)^t, \quad e_0\neq 0.
	\eeq 
	As it can be seen in Figure \ref{fig:E}, the effect space $\mathcal E$ is a closed convex double cone with a circular basis of radius $1/2$ located at height $e_0=1/2$ and vertices in $(0,0,0)$ and $(1,0,0)$, associated to the null and the unit effect, respectively \cite{BerthierProvenzi:2022PRS}. $\mathcal E$ represents the space of colors that can actually be sensed thanks to a perceptual measurement performed by an observer and its geometry happens to be in perfect agreement with that of the color solid advocated by Ostwald and de Valois, see e.g. \cite{Devalois:2000}.
	
	Effects parameterize a fundamental class of state transformations called \textit{Lüders operations}, which are \textit{convex-linear positive functions} $\psi_{\bf e}$ defined on the state space $\mathcal S$ and satisfying the constraint: 
	\beq
	0\le \Tr(\psi_{\bf e}(\rho))\le 1, \quad \text{for all } \rho\in \mathcal S.
	\eeq  
	After a Lüders operation, $\rho$ becomes a \textit{generalized density matrix}, whose set will be denoted with $\widetilde{\mathcal S}$, representing a \textit{post-measurement generalized state} which does not belong to the Bloch disk anymore, but to the cone generated by $\mathcal S$. The analytical expression of the post-measurement generalized state $\psi_ {\bf e}(  \rho )$, see  \cite{Busch:97}, is:
	\beq\label{eq:psies}
	\psi_ {\bf e}(\rho)=\eta_ {\bf e}^{1/2} \rho \eta_ {\bf e}^{1/2},
	\eeq 
	where $\eta_ {\bf e}^{1/2}$ is the Kraus operator associated to ${\bf e}$, the only symmetric and positive semi-definite matrix such that $\eta_ {\bf e}^{1/2}\eta_ {\bf e}^{1/2}=\eta_ {\bf e}$.
	
	Given a chromatic state ${\bf s}=(s_1,s_2)$ and the corresponding density matrix $\rho_{\bf s}:=\rho(s_1,s_2)$ of $\mathcal S$, we denote $\psi_ {\bf e}({\bf s}):=\psi_ {\bf e}( \rho_{\bf s})$.
	Thanks to the cyclic property of the trace,
	\beq\label{eq:expeff}
	\Tr(\psi_ {\bf e}(  {\bf s}))= \Tr(\rho_{\bf s} \, \eta_ {\bf e}) = \langle   {\bf e} \rangle_ {\bf s}=e_0(1+{\bf v}_{\bf e}\cdot {\bf v}_{\bf s}),
	\eeq
	and
	\beq\label{eq:varphi}   
	\varphi_ {\bf e}(  {\bf s}):=\frac{\psi_ {\bf e}(  {\bf s})}{\langle   {\bf e} \rangle_ {\bf s}}
	\eeq  
	is a density matrix representing to a genuine chromatic state belonging to $\mathcal S$.
	
	It is shown in \cite{BerthierProvenzi:2022PRS} that the state change $\bf s\mapsto \psi_{{\bf e}}({\bf s})$ can be computed with a 3-dimensional normalized Lorentz boost in the direction of $\bf v_e$, which implies
	\begin{equation}\label{eq:reladd2}
		\chi(\psi_{\bf e}({\bf s}))=e_0(1+{\bf v}_{\bf e}\cdot {\bf v}_{\bf s}) \, \frac{1}{2}\begin{pmatrix}1 \\{\bf v}_{\bf e}\oplus{\bf v}_{\bf s}\end{pmatrix} \in \widetilde{\mathcal S}.
	\end{equation}
	As a consequence the Bloch vector associated to the state $\varphi_{\bf e}({\bf s})$ is the vector
	\beq\label{eq:vphie}
	{\bf v}_{\varphi_{\bf e}({\bf s})} =   {\bf v_e} \oplus {\bf v_s},
	\eeq 
	where $\oplus$ denotes the Einstein-Poincaré relativistic sum. When $\|{\bf v_e}\|<1$, $\oplus$  is defined by
	\begin{equation}\label{eq:reladd}
		{\bf v_e}\oplus{\bf v_s}:=\frac 1{1+{\bf v_e}\cdot{\bf v_s}}\left\{{\bf v_e}+\frac1 {\gamma_{\bf v_e}}{\bf v_s}+\frac{\gamma_{\bf v_e}}{1+\gamma_{\bf v_e}}({\bf v_e}\cdot{\bf v_s}){\bf v_e}\right\},  
	\end{equation}
	where $\gamma_{\bf v_e}$ is the ${\bf e}$-Lorentz factor 
	\begin{equation}
		\gamma_{\bf v_e}:=\frac 1 {\sqrt{1-\Vert {\bf v_e}\Vert^2}},
	\end{equation}
	%\item if $\|{\bf v_e}\|=1$, then
	and, when $\|{\bf v_e}\|=1$, $\oplus$ is defined as follows
	\beq\label{eq:reladdunitvector}
	{\bf v_e}\oplus{\bf v_s}:={\bf v_e}.
	\eeq
	By convex-linearity, Lüders operations can be extended to generalized states as follows:
	\beq\label{eq:psigen}
	\psi_{\bf e}(s_0 {\bf s})= s_0 \psi_{\bf e}({\bf s}), \qquad \forall s_0\in [0,1],
	\eeq 
	this implies 
	\beq\label{eq:oneplus} 
	\langle{\bf e}\rangle_{s_0\bf s}=\Tr(s_0\rho_{\bf s}\eta_{\bf e})=s_0\Tr(\rho_{\bf s}\eta_{\bf e}) = s_0 \langle {\bf e} \rangle_{\bf s} = e_0 s_0 (1+{\bf v_e} \cdot \bf{v_s}),
	\eeq 
	so
	\beq\label{eq:intrchr}
	\varphi_{\bf e}(s_0{\bf s})=\frac{\psi_{\bf e}(s_0{\bf s})}{\langle {\bf e}\rangle_{s_0\bf s}} = \frac{s_0\psi_{\bf e}({\bf s})}{s_0\langle {\bf e}\rangle_{\bf s}} = \varphi_{\bf e}({\bf s}),
	\eeq 
	thus the post-measurement chromatic state depends solely on $\bf s$ and not on $s_0$. This implies the formula 
	\beq\label{eq:percareasplit}
	\psi_{\bf e}(s_0{\bf s}) =e_0 s_0 (1+{\bf v_e} \cdot \bf{v_s}) \, \varphi_{\bf e}({\bf s}),
	\eeq 
	which shows explicitly how the chromatic information about the state $\bf s$ and the expectation value of the effect $\bf e$ on $\bf s$ are \textit{fused together} in the post-measurement generalized state $\psi_{\bf e}(s_0{\bf s})$.
	
	In the case of an achromatic effect $\bf e_a$, for which ${\bf v_{e_a}}=\bf 0$, the previous formula gives
	\beq\label{eq:psiphiachr}
	\psi_{\bf e_a}(s_0{\bf s}) =e_0 s_0 \, \varphi_{\bf e_a}({\bf s}),
	\eeq 
	but $\eta_{\bf e_a}^{1/2}=\sqrt{e_0} \sigma_0$ so, by eq. \eqref{eq:psies}, 
	\beq\label{eq:psiachr}
	\psi_ {\bf e_a}(s_0{\bf s})=e_0s_0\rho_{\bf s},
	\eeq 
	hence $\varphi_{\bf e_a}({\bf s}) = \rho_{\bf s}$, or, by identifying $\rho_{\bf s}$ with the chromatic state $\bf s$,
	\beq\label{eq:achreffectstate} 
	\varphi_{\bf e_a}({\bf s}) = \bf s,
	\eeq 
	this means that the post-measurement state induced by the action of an achromatic effect coincides with the original state. 
	
	\subsection{The quantum information of perceived colors}\label{subsec:vocabrebit}
	According to the modeling described in the previous subsection, a formal framework for describing colorimetric perceptual attributes is provided by the following rules \cite{Berthier:22SIAM}:
	
	\begin{itemize}
		\item any quantity whose chromatic features manifest themselves fused together with a normalized scalar factor will be described through a \textit{generalized state};
		\item any act of (physical or perceptual) color measurement and the (physical or perceptual) medium used to perform it will be associated to an \textit{effect};
		\item the measurement outcome will be identified with the \textit{post-measurement generalized state} induced by the action of the effect.
	\end{itemize}
	The color stimulus hitting the eyes of the (human) observer can be either a light \textit{emitted} by a source of radiation or a light \textit{reflected} from the patch of a surface lit by an illuminant. 
	%Let us first consider the former situation.
	In the paper \cite{Berthier:22SIAM}, both situations were analyzed, however here we will concentrate only on emitted light because the experiments that we aim at performing will be conducted in the scenario of photons emitted by a source and sent to the eyes of the observer.
	
	The first set of definitions concerns the emitted light, the observer, and the color perceived by the observer from the emitted light.
	
	\bd[Emitted light stimulus] \label{def:emittedlight}
	An emitted light stimulus $\ell$ is identified with the generalized state $\ell_0 \bell$, $\ell_0\in [0,1]$ and $\bell \in \mathcal S$. The real quantity $\ell_0$ is the normalized light intensity and $\bell$ carries the intrinsic chromatic features.
	\ed 
	
	We denote with $\bf s_a$ the achromatic state represented by the density matrix $\rho_0$.
	
	\bd[Achromatic and white light] \label{def:achrlight}
	An achromatic light is an emitted light stimulus with $\ell_0\in [0,1]$ and $\bell = \bf s_a$. If, in particular, $\ell_0=1$, then we call it a white light and we write $\bell^W=\bf s_a$.
	\ed 
	
	Since a human observer is the medium through which a perceptual color measurement takes place, it is modeled as an effect.
	
	\bd[Observer]
	An observer $o$ measuring a color stimulus is identified with an effect ${\bf o}=(o_0,{\bf v_o}) \in \mathcal{E}$, $o_0\in [0,1]$ and ${\bf v_o} \in \cal D$. 
	\ed 
	
	The act of measuring an emitted light stimulus $\ell$ by an observer $o$ produces a perceived color through the Lüders operation associated to the effect $\bf o$.
	
	\bd[Perceived color from a light stimulus]\label{def:colorfromlight}
	Given an observer $o$ and an emitted light stimulus $\ell$, i.e. the couple $(\bf{o},\ell_0 \bell)$, the color perceived by $o$ from $\ell$ is the post-measurement generalized state $\psi_{\bf o}(\ell_0 \bell)\in \widetilde{\mathcal S}$.
	\ed 
	
	This definition is coherent with the three-dimensional nature of perceived colors, in fact eq. \eqref{eq:varphi} implies:
	\beq\label{eq:oexplicit} 
	\psi_{\bf o}(\ell_0 \bell) = o_0 \ell_0(1+{\bf v_o} \cdot {\bf v_{\bell}}) \, \varphi_{\bf o}(\bell) = \langle {\bf o} \rangle_{\ell_0 \bell} \  \varphi_{\bf o}(\bell), 
	\eeq 
	with $\langle {\bf o} \rangle_{\ell_0 \bell}\in  [0,1]$ and $\varphi_{\bf o}(\bell) \in \mathcal S$.
	
	Thanks to eq. \eqref{eq:achreffectstate}, we know that if an observer $o_{\text a}$ is associated to an achromatic effect $\bf o_a$, then $\varphi_{\bf o_a}({\bell}) = \bf \bell$,  which means that the chromatic state of the color perceived by $o_{\text a}$ from the light source $\ell =\ell_0 \bell$ is exactly its intrinsic chromatic state $\bell$.

	The second set of definitions concerns the achromatic perceptual attributes of perceived colors.
	%We can now pass to the definition of achromatic and chromatic attributes of perceived colors. The motivations underlying these definitions are given in \cite{Berthier:22SIAM}.
	
	\begin{definition}[Brightness of a perceived color from an emitted light]\label{def:brightlight}
		Given an observer $o$, ${\bf o}=(o_0,{\bf v_o})$, the brightness of the color $\psi_{\bf o}(\ell_0{\bell})$ perceived by $o$ from an emitted light stimulus $\ell_0 \bell$ is given by
		\begin{equation}\label{eq:brbr}
			{\mathcal B}(\psi_{\bf o}(\ell_0{\bell})):=\text{Tr}(\psi_{\bf o}(\ell_0{\bell}))=o_0\ell_0(1+{\bf v}_{\bf o}\cdot {\bf v}_{\bell}).
		\end{equation}
	\end{definition}
	Given any observer ${\bf o}=(o_0,{\bf v_o})$, the brightness perceived by $o$ from the white light $\bell^W$ is:
	\begin{equation}\label{eq:brightwlight}
		{\mathcal B}(\psi_{\bf o}(\bell^W))=o_0,
	\end{equation}
	so the brightness of the white light does not depend on the effect vector of $o$. 
	
	%\begin{definition}[Brightness of a perceived color from a reflected light]\label{def:brightrefl}
	% If $(o,\iota)$ represents a couple observer-illuminant, ${\bf o}=(o_0,{\bf v_o})$, ${\biota}=(\iota_0,{\bf v_{\biota}})$, the brightness of the color $\psi_{\bf o} (\psi_{\biota}(p_0{\bf p}))$ perceived by $o$ from a patch $p_o{\bf p}$ lit by $\iota$ is:
	%	\begin{equation}\label{eq:brightrefl}
		%		{\mathcal B}(\psi_{\bf o} (\psi_{\biota}(p_0{\bf p})) = o_0\iota_0p_0(1+{\bf v}_{{\biota}}\cdot{\bf v}_{\bf p})(1+{\bf v}_{{\bf o}}\cdot ({\bf v}_{{\biota}} \oplus {\bf v}_{\bf p})).
		%	\end{equation}
	%\end{definition}
	%The brightness perceived by $o$ from the white patch ${\bf p}^W$ lit by $\iota$ is:
	%	\begin{equation}\label{eq:brightwpatch}
		%		{\mathcal B}(\psi_{\bf o} (\psi_{\biota}({\bf p}^W)) = o_0\iota_0(1+{\bf v_o} \cdot {\bf v_{\biota}}),
		%	\end{equation}
	%	hence, the brightness of the white patch does not depend on the effect vector of $o$ if and only if $\iota$ is an achromatic illuminant $\iota_a$, in which case we have:
	%	\begin{equation}
		%		{\mathcal B}(\psi_{\bf o} (\psi_{\biota_{\bf p}}({\bf p}^W)) = o_0\iota_0.
		%	\end{equation}
	
	\begin{definition}[Lightness of a perceived color from an emitted light]\label{def:lightemitted}
		Given an observer $o$, ${\bf o}=(o_0,{\bf v_o})$, the lightness of the color $\psi_{\bf o}(\ell_0{\bell})$ perceived by $o$ from an emitted light stimulus $\ell_0 \bell$ is given by the ratio between its brightness, eq. \eqref{eq:brbr}, and the brightness of the white light, eq. \eqref{eq:brightwlight}, i.e.
		\begin{equation}
			{\mathcal L}(\psi_{\bf o}(\ell_0{\bell})):=\frac{{\mathcal B}(\psi_{\bf o}(\ell_0{\bell}))}{{\mathcal B}(\psi_{\bf o}({\bell}^W))}=\ell_0(1+{\bf v}_{\bf o}\cdot {\bf v}_{\bell}).
		\end{equation}
		
	\end{definition}
	The lightness perceived from an achromatic emitted light coincides with its normalized intensity $\ell_0$ independently of the observer:
	\beq
	{\mathcal L}(\psi_{\bf o} (\ell_0{\bf s_a}))=\ell_0, \quad \forall {\bf o}.
	\eeq 
	In particular, the lightness of the white light is normalized to 1. 
	
	When ${\bf v}_{\bf o}=\bf 0$, the lightness of any color perceived from an emitted light coincides with the light intensity independently of the chromatic state of the emitted light:
	\beq
	{\mathcal L}(\psi_{\bf o_a} (\ell_0\bell))=\ell_0, \quad \forall \bell.
	\eeq

	Finally, the last set of definitions concerns the chromatic perceptual attributes of perceived colors.
	Given the perceived color $\psi_{\bf o}({\bf s})$, its saturation is
	\begin{equation}\label{eq:saturation}
		\begin{split}
			{\rm Sat}(\psi_{\bf o}(\rho ))& =R(\rho_{\varphi_{\bf o}(\rho )}||\rho_{\bf 0})\\
			& = \frac 1 2 \log_2(1-r_{\varphi_{\bf o}(\rho )}^2)+\frac{r_{\varphi_{\bf o}(\rho )}} 2\log_2\left(\frac{1+r_{\varphi_{\bf o}(\rho )}}{1-r_{\varphi_{\bf o}(\rho )}}\right),
		\end{split}
	\end{equation}
	where $R$ is the \textit{relative entropy}\footnote{we recall that $R(\rho_{{\bf s}}||\rho_{{\bf t}}):=\Tr\left[ \rho_{{\bf s}}\log_2\rho_{{\bf s}}-\rho_{{\bf s}}\log_2\rho_{{\bf t}}\right]$.} between the states appearing as its arguments and $r_{\varphi_{\bf o}(\rho)}=\|{\bf v}_{\varphi_{\bf o}(\rho)}\|$, while its hue is the pure chromatic state $\varphi^*_{\bf o}(\rho)$ defined by
	\begin{equation}\label{eq:hue}
		\varphi^*_{\bf o}(\rho):=\underset{\rho\in \cal PS }{\arg\min} \; R(\rho||\rho_{\varphi_{\bf o}(\rho)}),
	\end{equation}
	where $\cal PS$ are the pure chromatic states, parameterized by the points of the border $\partial \cal D$ of the Bloch disk. The explicit expression of the density matrix associated to the pure chromatic state $\varphi^*_{\bf o}(\rho)$ is 
	\begin{equation}\label{hue}
		\rho_{\varphi^*_{\bf o}(\rho )}=\frac 1 2\left(\begin{array}{cc}1+\cos\vartheta_{\varphi_{\bf o}(\rho )} & \sin\vartheta_{\varphi_{\bf o}(\rho )} \\\sin\vartheta_{\varphi_{\bf o}(\rho )} & 1-\cos\vartheta_{\varphi_{\bf o}(\rho )}\end{array}\right),
	\end{equation}
	where
	\beq 
	\cos\vartheta_{\rho,\varphi_{\bf o}(\rho )}=\frac{{\bf v}_{\rho} \cdot {\bf v}_{\varphi_{\bf o}(\rho )}}{r_{\varphi_{\bf o}(\rho )}}.
	\eeq

	\section{Perceived colors from the qubit point of view}
	\label{sec:recast}
	The aim of this section is to show how to compute the perceptual attributes of perceived colors in the more usual physical context where the Hilbert space of the quantum system is the complex vector space $\mathbb C^2$ rather than the real vector space $\mathbb R^2$ of the rebit system.
	
	\subsection{From rebit generalized states and effects to qubit states and effects}
	
	The set $\mathcal H(2,\C)$ of $2\times 2$ hermitian matrices is a $\R$-vector space of dimension 4 with an orthogonal basis 
	with respect to the Hilbert-Schmidt inner product $\langle A,B\rangle_{\text{HS}} := \Tr(A^\dag B)$
	given by the Bloch basis $(\sigma_0,\sigma_x,\sigma_y,\sigma_z)$, where $\sigma_0$ is again $I_2$ and 
	\beq 
	\sigma_x:=\pmat{0 & 1 \\ 1 & 0}, \quad \sigma_y:=\pmat{0 & -i \\ i & 0}, \quad \sigma_z:=\pmat{1 & 0 \\ 0 & -1},
	\eeq 
	constitute the full set of Pauli matrices.
	The squared Hilbert-Schmidt norm of all the elements of the Bloch basis is 2, so a density matrix belonging to $\mathcal H(2,\C)$ can be written as follows:
	\beq\label{eq:rhoxyz}
	\rho(x,y,z)=\frac{1}{2}(I_2+x \sigma_x+y \sigma_y+z \sigma_z)=
	\frac{1}{2}
	\pmat{
		1+z&x-i y\\
		x+i y&1-z},
	\eeq
	where $x^2+y^2+z^2\le 1$, a condition that defines the \textit{Bloch sphere}. As it is well-known, a quantum system whose states can be described by Bloch vectors belonging to this sphere is a \textit{qubit}. The expression of a Bloch vector in spherical coordinates is 
	\beq
	{\bf v}=r\, \pmat{\sin\theta\cos\phi \\ \sin\theta\sin\phi \\ \cos\theta},
	\eeq 
	with $r\in [0,1]$, $\theta\in [0,\pi]$ and $\phi\in [0,2\pi)$, and the associated density matrix is 
	\beq \rho(r,\theta,\phi)=\frac{1}{2}\pmat{
		1+r \cos\theta & e^{-i \phi} r \sin\theta\\
		e^{i \phi} r \sin\theta &1-r \cos\theta
	}.
	\eeq 
	
	For the sake of a more concise writing, from now on we will indicate with `r-' and `q-' the quantities relative to the rebit and qubit formulations, respectively.
	
	First, we discuss how to define an r-emitted light as a q-state. 
	In Def. \ref{def:emittedlight}, a r-emitted light stimulus is written as the r-generalized state $\ell_0 \bell$, with $\ell_0$ representing the normalized intensity and $\bell\in\mathcal{S}$ the intrinsic chromatic state. In the r-framework, we need a generalized state in order to take into account the information of both $\ell_0$ and $\bell$, however we are going to show that in the q-formalism a complex density matrix can carry the entire information. 
	
	The advantage is that this density matrix can be more easily associated to the actual preparation of a photon to be sent in the eyes of an observer, as will become explicit in Definition \ref{def:qlightstimulus}, together with the accompanying discussion, where we formalize how such a density matrix encodes the chromatic state of an emitted light stimulus in a form directly applicable to an experimental setup.		
	
	We are thus led to define a map from the space $\widetilde {\mathcal S}$ of r-generalized states to the Bloch sphere of q-states. As mentioned earlier, $\ell_0=0$ represents the \textit{threshold of photopic visibility}, i.e. the minimal light intensity that can induce the response of a cone in the retina, while $\ell_0=1$ represents the \textit{glare limit}, when all light stimuli are perceived as the brightest possible white and, above that intensity, vision is no longer possible because the cones saturate. 
	
\begin{figure}[!ht]
		\centering
		\includegraphics[width=5 cm]{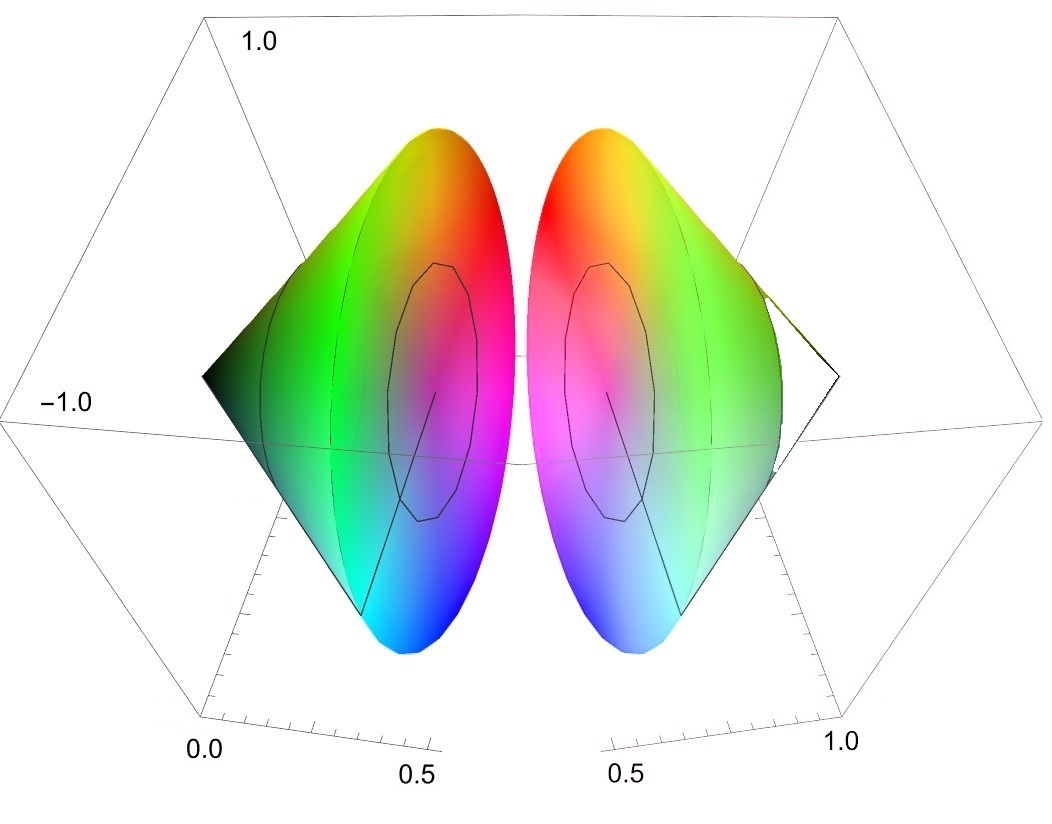}
		\caption{Geometric representation of the set $\mathcal E_1$ opened in half at its basis.\label{fig2}}
	\end{figure}

\begin{figure}[!ht]
		\centering
		\includegraphics[width=10.5 cm]{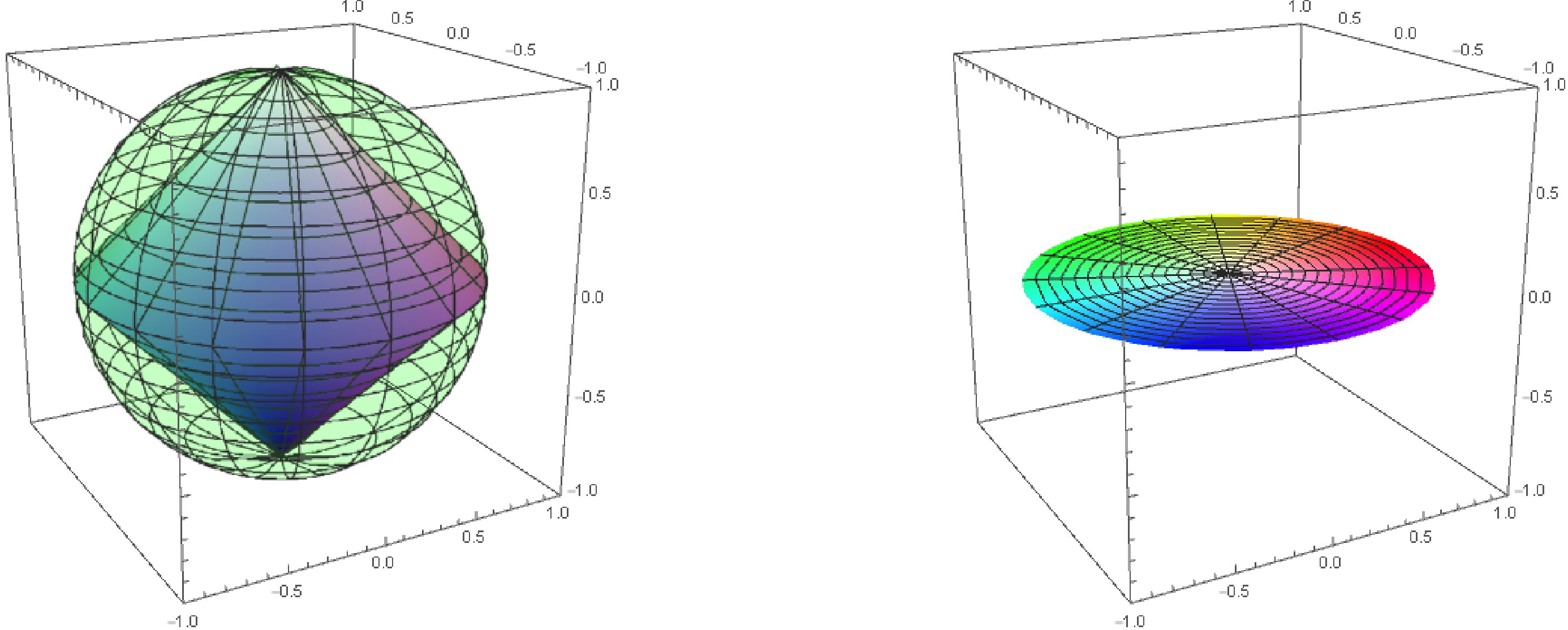}
		\caption{\textit{Left}: geometric representation of the set $\mathcal E_2$ inside the Bloch sphere. \textit{Right}: the equatorial basis of the double cone corresponds to the Bloch disk with Hering's opponency.\label{fig3}}
	\end{figure}

	This means that the only \textit{meaningful light stimuli} are those belonging to the subset of $\widetilde {\mathcal S}$ given by all $\psi_{\bf e}(\bf s_a)$, as $\bf e$ varies in $\mathcal E$, i.e.
	\beq
	\mathcal{E}_1:=\left\{(\ell_0,\ell_1,\ell_2)\in \R^3, \, \ell_0\in[0,1] \text{ and } (\ell_0 \ell_1)^2+(\ell_0 \ell_2)^2\leq  \min\{\ell_0^2,(1-\ell_0)^2\} \right\},
	\eeq 
	see Figure \ref{fig2}. It is convenient to represent the light stimuli belonging to $\mathcal{E}_1$ within the Bloch sphere since it contains the q-states. This can be achieved thanks to the following bijection
	\beq\label{eq:bijection}
	\begin{array}{cccl}
		\beta: & \mathcal{E}_1\subset\widetilde{\mathcal S}  & \longrightarrow & \mathcal{E}_2   \\
		& (\ell_0,\ell_1,\ell_2) & \longmapsto         & \beta(\ell_0,\ell_1,\ell_2):=(\ell_0 \ell_1,\ell_0 \ell_2,2 \ell_0-1),
	\end{array}
	\eeq 
	where $\mathcal E_2$ is the subset of the Bloch sphere defined by
	\beq
	\mathcal{E}_2:=\left\{(x,y,z)\in \R^3, \, z\in[-1,1] \text{ and } x^2+y^2\leq (1-|z|)^2 \right\}.
	\eeq 
	$\mathcal E_2$ is a solid convex double cone, with vertices in $(0,0,1)$ and $(0,0,-1)$. Its circular basis is now of radius $1$, located at height $z=0$, and it can be interpreted as the Bloch disk of the r-framework, as depicted in Figure \ref{fig3}.
	
	So the q-density matrix associated to the r-emitted light stimulus $\ell_0\bell$ is 
	\beq\label{eq:rhobell}
	\rho_{\bell} =\frac 1 2\left( I_2 + \ell_0\ell_1 \sigma_x + \ell_0\ell_2\sigma_y + (2\ell_0-1)\sigma_z \right) = \frac 1 2 \pmat{2\ell_0 &  \ell_0\ell_1 - i \ell_0\ell_2 \\ \ell_0\ell_1 + i \ell_0\ell_2 & 2(1-\ell_0)},
	\eeq
	with q-Bloch vector given by
	\beq\label{eq:Blochvecqdensmat}
	{\bf v}^q_{\bell}=\pmat{\ell_0\ell_1 \\ \ell_0\ell_2 \\2\ell_0-1}.
	\eeq 
	It is possible to decompose $\rho_{\bell}$ as follows
	\beq
	\rho_{\bell} = p_w\pmat{
		1&0\\
		0&0} +
	p_b \pmat{
		0&0\\
		0&1} +
	\frac{p_c}{2}\pmat{
		1&s e^{-i\phi}\\
		s e^{i\phi}&1}
	\equiv p_w\rho_w+p_b\rho_b+p_c \rho_c,
	\eeq 
	with
	%In $\rho_{\bell}$ the information about the light stimulus is separated into its diagonal and off-diagonal parts: the former carries only the information about the normalized intensity $\ell_0$, while the latter contains the chromatic information and $\ell_0$ appears as an overall factor. 
	%In particular, thanks to Def. \ref{def:achrlight}, if $\ell_1=\ell_2=0$ and $\ell_0=1$, then we get a white light, while $\ell_0=0$ corresponds to the absence of light, or `black'.
	%These considerations inspire the curiosity to investigate the possibility to write $\rho_{\bell}$ as a convex combination of three meaningful q-density matrices, two of which representing  white light and black and the third representing the chromatic component of $\rho_{\bell}$.
	%This task can be achieved by considering three coefficients $p_w,p_b,p_c\in [0,1]$, where clearly the letters $w,b,c$ stand for white, black and chromatic, respectively, and imposing
	\beq\label{eq:ps}   
	\begin{cases}
		2\ell_0=2 p_w+p_c \\
		2(1-\ell_0)=2 p_b+p_c \\
		\ell_0\ell_1\pm i\ell_0\ell_2=p_c\, s e^{\pm i\phi},
	\end{cases}
	\eeq 
	and where $s\in [0,1]$, $\phi\in [0,2\pi)$, and $p_w,p_b,p_c\in [0,1]$. 
	
	The condition $\ell_1=\ell_2=0$ defines an achromatic r-emitted light. In particular, if $\ell_0=1$ or $\ell_0=0$, we have a white or black r-emitted light, indicated with $\rho_w$ or $\rho_b$, respectively. Instead, $\rho_c$, which describes the chromatic component of the r-emitted light $\ell_0\bell$, is characterized by $\ell_0=1/2$ and $\ell_1$ and/or $\ell_2$ non-null.
	
	The concept of observer in the q-framework, which amounts at defining the q-analogue of an r-effect $\bf e\in \mathcal E$. This can be done by considering the matrix 
	\beq\label{eq:effmat}
	\rho_{\bf e}:=\frac 1 2 \pmat{1+e_z & e_x-ie_y \\ e_x+ie_y & 1-e_z},
	\eeq 
	with 
	\beq
	{\bf v}^q_{\bf e}:=\pmat{e_x\\ e_y \\ e_z} = \pmat{e_0 e_1 \\ e_0e_2 \\ 2e_0-1},
	\eeq 
	$e_z\in[-1,1]$, and imposing the condition ${\bf 0}\le \rho_{\bf e} \le I_2$.  By direct computation, it can be verified that this is equivalent to the condition $0\le e_x^2+e_y^2+e_z^2\le 1$, i.e. that ${\bf v}^q_{\bf e}$ belongs to the Bloch sphere. Hence, the vector ${\bf v}^q_{\bf e}$ serves simultaneously as Bloch and effect vector for the matrix $\rho_{\bf e}$.
	
	It can be checked that the resolution of the identity can be written as $I_2=\rho_{\bf e}+\rho_{-{\bf e}}$, so we can limit the variability of the parameters defining $\rho_{\bf e}$ to the upper half Bloch sphere, i.e. $e_z\in [0,1]$ and $e_x^2+e_y^2+e_z^2\le 1$, and reconstruct the information from the lower half Bloch sphere via the antipodal q-vectors $-{\bf v}^q_{\bf e}=-(e_x,e_y,e_z)$. 
	
	We adopt the following definition.
	\bd[q-effect]
	A q-effect is a q-state 
	\beq\label{eq:queff}     
	\rho_{\bf e}=\frac 1 2 \pmat{1+e_z & e_x-ie_y \\ e_x+ie_y & 1-e_z},
	\eeq 
	with $e_z\in [0,1]$ and $e_x^2+e_y^2+e_z^2\le 1$.
	\ed 
	As it is well-known, extremal effects are orthogonal projectors and they are labeled by unit vectors in the Bloch sphere. As an example, the Bloch-effect vector with $e_0=1$, $e_1=e_2=0$, i.e. ${\bf v}^q_{\bf e}=(0,0,1)^t$ labels the orthogonal projector on the `white point'
	\beq
	P_w:=\pmat{1 & 0 \\ 0 & 0},
	\eeq 
	while the one identified by $e_0=e_1=e_2=0$, which coincides with its antipodal vector $-{\bf v}^q_{\bf e}=(0,0,-1)^t$ labels the orthogonal projector on the `black point'
	\beq
	P_b:=\pmat{0 & 0 \\ 0 & 1},
	\eeq 
	and together they form a resolution of the identity: $P_w+P_b=I_2$.
	
	\subsection{Color measurements with qubit states and effects}
	
	Here, using eqs. \eqref{eq:rhobell}, \eqref{eq:ps}, \eqref{eq:queff}, we formalize the definitions that permit to describe perceived colors as results of measurements using qubit states and effects, rather than their rebit counterparts.
	\bd[q-emitted light stimulus]\label{def:qlightstimulus}
	A q-emitted light stimulus $\ell$ is identified with the following q-density matrix:
	\beq\label{eq:rhoell}
	\rho_{\bell}=\dfrac{1}{2}\pmat{
		2p_w+p_c & p_c\, s\, e^{-i \phi}\\
		p_c\, s\, e^{i \phi}&2p_b+p_c
	},
	\eeq 
	where $p_b,p_w,p_c,s\in[0,1]$ and $p_b+p_w+p_c=1$.
	\ed
	In particular:
	\bi
	\item a q-achromatic light is an emitted light stimulus with $p_c=0$;
	\item the q-white light $\rho_W$ is a q-achromatic light with $p_w=1$.
	\ei 
	Hence, the generic q-achromatic emitted light can be written as 
	$$\rho^a_{\bell}=\frac 1 2\pmat{
		1+z & 0\\
		0 & 1-z
	}, \quad z\in[-1,1].$$
	As noted at the beginning of this section, the formulation of a q-emitted light stimulus in Definition \ref{def:qlightstimulus} makes the distinction between chromatic and achromatic light completely explicit. In particular, the chromatic component of $\rho_{\bell}$ is encoded by the parameters $p_c$, $s$, and $\phi$, while the achromatic component is entirely contained in the diagonal entries via $p_w$ and $p_b$.
	
	This separation of roles within the density matrix not only mirrors the conceptual division between chromatic and achromatic components of the light stimulus, but also provides a form directly suited to experimental preparation in laboratory settings.
	
	\bd[q-observer]\label{def:qobserver}
	A q-observer $o$ measuring a color stimulus is identified with a q-state 
	\beq
	\rho_{\bf o}=\frac 1 2 \pmat{1+o_z & o_x-io_y \\ o_x+io_y & 1-o_z}=\frac{1}{2}\pmat{
		1+r_o \cos \theta_o&e^{-i \phi_o} r_o \sin \theta_o\\
		e^{i \phi_o} r_o \sin \theta_o&1-r_o \cos \theta_o
	},
	\eeq 
	with $o_z,r_o\in [0,1]$, $o_x^2+o_y^2+o_z^2\le 1$, and\footnote{taking $\theta_o\in[0,\frac{\pi}{2}]$ is equivalent to considering only the upper half cap of the Bloch sphere.} $\theta_o\in[0,\frac{\pi}{2}]$, $\phi_o\in[0,2\pi)$.
	\ed

	In the r-framework, we obtain a perceived color from an emitted light stimulus $\ell_0\bell$ by applying on it the Lüders operation $\psi_{\bf o}$ related to the effect that represents the observer $o$. The result is the r-generalized state $\psi_{\bf o}(\ell_0\bell)$. 
	
	By eq. \eqref{eq:rhobell}, the r-generalized state $\ell_0\bell$ is represented by the q-density matrix $\rho_{\bell}$ in the q-framework. Coherently with that, also the perceived color given by the r-generalized state $\psi_{\bf o}(\ell_0\bell)$, has to be represented in the q-framework by a q-density matrix that we write $\rho_{\bell}^o:=\varphi_{\bf o}(\rho_{\bell})$ and we call post-measurement q-state. 
	
	If we denote with $c_0,c_1,c_2$ the matrix element of $\rho_{\bell}^o$, then by eq. \eqref{eq:rhobell} we have 
	\beq
	\rho_{\bell}^o = \frac 1 2 \pmat{2 c_0 & c_0(c_1-ic_2) \\ c_0(c_1+ic_2) & 2(1-c_0)},
	\eeq
	%=\frac{1}{2}\pmat{
		%	1+r_c \cos \theta_c&e^{-i \phi_c} r_c \sin \theta_c\\
		%	e^{i \phi_c} r_c \sin \theta_c&1-r_c \cos \theta_c
		and, by eq. \eqref{eq:Blochvecqdensmat}, the relative Bloch vector is
		\beq
		{\bf v}^o_{\bell} := \pmat{c_0 c_1 \\ c_0 c_2 \\ 2 c_0-1}.
		\eeq 
		%= r_c \pmat{\cos \phi_c\sin \theta_c \\ \sin \phi_c\sin \theta_c \\ \cos \theta_c},
		In order to find the explicit expression of the entries $c_0,c_1,c_2$, let us introduce the following `chromatic q-Bloch vectors':  
		\beq\label{eq:chromaticqvectors}
		\overline{{\bf v}}^q_{\bell}:=\pmat{\ell_1 \\ \ell_2}, \quad \overline{{\bf v}}^q_{\bf o}:=\pmat{{o_x}/{o_z}\\ {o_y}/{o_z}}, \quad \overline{{\bf v}}^o_{\bell} := \pmat{c_1 \\ c_2},
		\eeq 
		where the presence of $o_z$ in $\overline{{\bf v}}^q_{\bf o}$ is reminiscent of the presence of $e_0$ in eq. \eqref{eq:veceff}. Using equations from \eqref{eq:reladd2} to \eqref{eq:reladdunitvector} and eq. \eqref{eq:oexplicit}, with straightforward computations we find
		\bi 
		\item if $\|\overline{{\bf v}}^q_{\bf o}\|<1$, then 
		\beq
		\begin{cases}
			c_0 = o_z \ell_0 (1+ \overline{{\bf v}}^q_{\bf o}\cdot \overline{{\bf v}}^q_{\bell})\\
			\overline{{\bf v}}^o_{\bell} =  \dfrac{1}{1+\overline{{\bf v}}^q_{\bf o}\cdot \overline{{\bf v}}^q_{\bell}}\left\{\overline{{\bf v}}^q_{\bf o}+\dfrac{1}{\gamma_{\bf o}}\overline{{\bf v}}^q_{\bell} + \dfrac{\gamma_{\bf o}}{1+\gamma_{\bf o}}(\overline{{\bf v}}^q_{\bf o}\cdot \overline{{\bf v}}^q_{\bell}) \overline{{\bf v}}^q_{\bf o}\right\} 
		\end{cases},
		\eeq 
		where 
		\beq\label{eq:gammao}
		\gamma_{\bf o}:=1/{\sqrt{1-\|\overline{{\bf v}}^q_{\bf o}\|^2}};
		\eeq 
		\item if $\|\overline{{\bf v}}^q_{\bf o}\|=1$, then %, since $o_z^2+\|\overline{{\bf v}}^q_{\bf o}\|^2\le 1$, $o_z$ must be null, so
		\beq
		\begin{cases}
			c_0 = o_z \ell_0 (1+ \overline{{\bf v}}^q_{\bf o}\cdot \overline{{\bf v}}^q_{\bell})\\
			\overline{{\bf v}}^o_{\bell} \underset{\text{eq. } \eqref{eq:reladdunitvector}}{=}  \overline{{\bf v}}^q_{\bf o}
		\end{cases}.
		\eeq  
		\ei 
		It follows that the definition of perceived color from an emitted light stimulus can be formalized as follows.
		\begin{definition}[q-perceived color from an emitted light stimulus]
			Given the q-effect $\rho_{\bf o}$ as in Def. \ref{def:qobserver}  associated to the observer $o$ and an the emitted light stimulus $\ell$ described by the q-state $\rho_{\bell}$ as in eq. \eqref{eq:rhobell}, the color perceived by $o$ from $\ell$ is the post-measurement q-state
			\beq \rho_{\bell}^o=\frac 1 2 \pmat{2 c_0 & c_0(c_1-ic_2) \\ c_0(c_1+ic_2) & 2(1-c_0)}
			\eeq 
			%	=\frac{1}{2}\pmat{
				%		1+r_c \cos \theta_c&e^{-i \phi_c} r_c \sin \theta_c\\
				%		e^{i \phi_c} r_c \sin \theta_c&1-r_c \cos \theta_c},
			where
			\bi 
			\item $c_0= o_z \ell_0 (1+ \overline{{\bf v}}^q_{\bf o}\cdot \overline{{\bf v}}^q_{\bell})$, with $\overline{{\bf v}}^q_{\bf o}$ and $\overline{{\bf v}}^q_{\bell}$ as in eq. \eqref{eq:chromaticqvectors};
			\item and $\overline{{\bf v}}^o_{\bell} = \overline{{\bf v}}^q_{\bf o} \oplus \overline{{\bf v}}^q_{\bell}$, i.e. 
			\beq 
			\overline{{\bf v}}^o_{\bell}=\pmat{c_1\\ c_2}=
			\begin{cases}
				\overline{{\bf v}}^q_{\bf o} & \text{ if } \|\overline{{\bf v}}^q_{\bf o}\|=1\\
				\dfrac{1}{1+\overline{{\bf v}}^q_{\bf o}\cdot \overline{{\bf v}}^q_{\bell}}\left\{\overline{{\bf v}}^q_{\bf o}+\dfrac{1}{\gamma_{\bf o}}\overline{{\bf v}}^q_{\bell} + \dfrac{\gamma_{\bf o}}{1+\gamma_{\bf o}}(\overline{{\bf v}}^q_{\bf o}\cdot \overline{{\bf v}}^q_{\bell}) \overline{{\bf v}}^q_{\bf o}\right\} & \text{ otherwise}
			\end{cases},
			\eeq 
			with $\gamma_{\bf o}$ as in eq. \eqref{eq:gammao}.
			\ei 
		\end{definition}
		It can be seen that the q-chromatic vector $\overline{{\bf v}}^o_{\bell}$ coincides with the vector 
		${\bf v}_{\varphi_{\bf e}({\bf s})}$ in eq. \eqref{eq:vphie} defining the post-measurement chromatic state in the r-framework. 
		
		\subsection{Computing perceptual attributes with qubit states and effects}
		
		It can be seen that $c_0$ has the same expression of the brightness of a light stimulus defined in eq. \eqref{eq:brbr} which may seem odd because the brightness was obtained in the r-framework by taking the trace of the post-measurement generalized states $\psi_{\bf o}(\ell_0\bell)$. However, in the q-formalism we cannot hope to obtain the brightness through the trace of $\rho_{\bell}^o$, because this quantity is always equal to 1.
		
		For this reason, we have to come up with a novel and meaningful way to compute the brightness in the q-formalism which has to be coherent with the result obtained in the r-framework. To this aim, we observe that the $z$-axis of the Bloch sphere depicted in Figure \ref{fig3} contains \textit{non-normalized} achromatic values. Moreover, the product of $\rho_{\bell}^o$ with the Pauli matrix $\sigma_z$ produces the $z$-component of the Bloch vector of $\rho_{\bell}^o$. 
		
		So, it makes sense to compute the non-normalized brightness of the perceived color $\rho_{\bell}^o$ as the expectation value of $\sigma_z$ on the state defined by $\rho_{\bell}^o$, that is
		\beq
		\Tr(\rho_{\bell}^o \, \sigma_z) = 2 c_0 -1.
		\eeq 
		Since the $z$-component in the Bloch sphere ranges from $-1$ to $+1$, to obtain the brightness of $\rho_{\bell}^o$ we need to add 1 and divide by 2 in order to arrive to an achromatic value between 0 and 1. This explains why in the q-formalism the definition of brightness coherent with the r-framework is the following.
		
		\begin{definition}[q-brightness]
			Given a perceived color $\rho_{\bell}^o$ from a light stimulus $\rho_{\bell}$, its brightness is defined as follows
			\beq \mathcal B(\rho_{\bell}^o):=\frac{1+\emph{Tr}(\rho_{\bell}^o\,\sigma_z)}{2}=c_0=o_z \ell_0+o_x \ell_0 \ell_1+o_y \ell_0 \ell_2.\eeq 
		\end{definition}
		In particular, for the q-white light $\rho_W$ we have $\ell_0=1$ and $\ell_1=\ell_2=0$, so 
		\beq
		\mathcal B(\rho_{W}^o)=o_z.
		\eeq  
		As a consequence, in order to be coherent with the r-framework, the q-lightness must be defined dividing the brightness expression by $o_z$. However, $o_z$ is the expectation value of $\sigma_z$ on the state $\rho_{\bf o}$, thanks to the already underlined duality state-effect in the q-formalism. 
		
		These remarks lead naturally to the following definition.
		\begin{definition}[q-lightness]
			Given a perceived color $\rho_{\bell}^o$ from a light stimulus $\rho_{\bell}$, its lightness is defined as follows
			$$\mathcal L(\rho_{\bell}^o):=\frac{1+\emph{Tr}(\rho_{\bell}^o\,\sigma_z)}{2 \emph{Tr}(\rho_{\bf o}\,\sigma_z)}=\ell_0+\frac{o_x}{o_z} \ell_0 \ell_1+\frac{o_y}{o_z} \ell_0 \ell_2,$$
			where $\rho_{\bf o}=\rho(o_x,o_y,o_z)$ is the q-effect associated with the observer $o$.
		\end{definition}
		In the r-framework, the saturation of a perceived color defined as in eq. \eqref{eq:saturation} provides a  measure of how the post-measurement chromatic state can be discerned from the achromatic one using as measure of distinguishability the relative quantum entropy.
		
		To translate eq. \eqref{eq:saturation} in the q-framework we need the Euclidean norm of the Bloch vector ${\bf v}^o_{\bell}=(c_1,c_2)^t$, which is 
		\beq
		r_{\rho_{\bell}^o}=\sqrt{c_1^2+c_2^2},
		\eeq 
		where $c_1$ and $c_2$ can be obtained via $\rho_{\bell}^o$ as follows:
		\beq
		c_1=\frac{\Tr(\rho_{\bell}^o \, \sigma_x)} {c_0}=\frac{2\Tr(\rho_{\bell}^o \, \sigma_x)}{1+\Tr(\rho_{\bell}^o \, \sigma_z)}, \quad c_2=\frac{\Tr(\rho_{\bell}^o \, \sigma_y)} {c_0}=\frac{2\Tr(\rho_{\bell}^o \, \sigma_y)}{1+\Tr(\rho_{\bell}^o \, \sigma_z)},
		\eeq 
		so that 
		\beq\label{eq:radius}
		r_{\rho_{\bell}^o}=\frac{2}{1+\Tr(\rho_{\bell}^o\,\sigma_z)}\sqrt{\Tr(\rho_{\bell}^o\,\sigma_x)^2+\Tr(\rho_{\bell}^o\,\sigma_y)^2},
		\eeq 
		with $\Tr(\rho_{\bell}^o\sigma_z)>-1$.
		
		\begin{definition}[q-saturation]
			Given a perceived color $\rho_{\bell}^o$ from a light stimulus $\rho_{\bell}$, its saturation is defined as follows
			\beq {\rm Sat}(\rho_{\bell}^o):=\frac 1 2\log_2(1-r_{\rho_{\bell}^o}^2)+\frac{r_{\rho_{\bell}^o}} 2\log_2\left(\frac{1+r_{\rho_{\bell}^o}}{1-r_{\rho_{\bell}^o}}\right),
			\eeq 
			with $r_{\rho_{\bell}^o}$ as in eq. \eqref{eq:radius}.
		\end{definition}
		
		Finally, let us discuss the hue. In the r-framework, the hue of a perceived color, see eq. \eqref{eq:hue}, is the pure state that minimizes the relative entropy with the post-measurement state. As shown by eq. \eqref{hue}, in its definition there appear the polar angle of the Bloch vector associated to the post-measurement state. 
		
		In the q-framework, the analogous of that angle is given by 
		\beq\label{eq:phi}
		\phi=2 \arctan \frac{c_2}{\sqrt{c_1^2+c_2^2}+c_1}=2 \arctan \frac{\Tr(\rho_{\bell}^o\,\sigma_y)}{\sqrt{\Tr(\rho_{\bell}^o\,\sigma_x)^2+\Tr(\rho_{\bell}^o\,\sigma_y)^2}+\Tr(\rho_{\bell}^o\,\sigma_x)},
		\eeq 
		with at least one between the values $\Tr(\rho_{\bell}^o\,\sigma_x),  \Tr(\rho_{\bell}^o\,\sigma_y)$ strictly larger than 0. 
		
		\begin{definition}[q-hue]
			Given a perceived color $\rho_{\bell}^o$ from a light stimulus $\rho_{\bell}$, its hue is defined as the following pure chromatic state 
			$$\rho^o_{\bell,\phi}:=
			\frac{1}{2}\pmat{
				1& e^{-i\phi}\\
				e^{i\phi}&1},$$
			with $\phi$ as in eq. \eqref{eq:phi}.
		\end{definition}

		\subsection{The operational framework}
		%\section{The structure of the q-formalism for color perception}
		\label{sec:PauliBloch}
		%In this section provide an abstract formalization of the set of density operators on the Hilbert space $\C^2$ that we have defined so-far which serve as structure for the q-formalism. This structure is represented by:
		The operational framework that we propose to perform experiments regarding the quantum model of color perception is defined by the following data:
		$$(\mathcal {\bf \Lambda},\xi, \sigma_x,\sigma_y,\sigma_z,\phi_\text{white},\phi_\text{red},\phi_\text{green},\phi_\text{yellow},\phi_\text{blue}),$$
		where 
		\bi 
		\item ${\bf \Lambda}$ is the set of all q-emitted light stimuli 
		\beq {\bf \Lambda}=\left\{\rho=\pmat{
			p_w+\frac{p_c}{2}&\frac{p_c\, s\, e^{-i \phi}}{2}\\
			\frac{p_c\, s\, e^{i \phi}}{2}&p_b+\frac{p_c}{2}
		} : p_b,p_w,p_c,s\in[0,1] \text{ and }p_b+p_w+p_c=1\right\} \, ;
		\eeq
		\item $\xi$ is the set of q-effects describing q-observers who perform perceptual measurements on ${\bf \Lambda}$ giving rise to post-measurement q-states corresponding to perceived colors
		\beq
		\xi =\left\{\frac{1}{2}\pmat{
			1+r \cos\theta&e^{-i \phi} r \sin\theta\\
			e^{i \phi} r \sin\theta&1-r \cos\theta
		}: r\in [0,1], \theta \in [0,\pi/2], \phi \in [0,2\pi) \right\} \, ;
		\eeq
		\item the Pauli matrices $\sigma_x,\sigma_y,\sigma_z$ can be interpreted dually as extremal effects or pure states (rank-1 projectors) given by
		\beq
		\sigma_x=\xi_{1,\frac{\pi}{2},0}, \quad \sigma_y=\xi_{1,0,\frac{\pi}{2}}, \quad  \sigma_z=\xi_{1,0,0} \, ;
		\eeq 
		\item the \textit{constants} $\phi_\text{white},\phi_\text{red},\phi_\text{green},\phi_\text{yellow},\phi_\text{blue}$ are pure states associated to the pure white and the antipodal hues:
		\beq \phi_\text{white}=\pmat{
			1&0\\
			0&0
		},
		\eeq
		\beq
		\phi_\text{red}=\frac{1}{2}\pmat{
			1&1\\
			1&1
		}, \quad \phi_\text{green} = \sigma_z \, \phi_\text{red} \,  \sigma_z, 
		\eeq
		\beq 
		\phi_\text{yellow}=\frac{1}{2}\pmat{
			1&-i\\
			i&1
		}, \quad \phi_\text{blue} = \sigma_z \, \phi_\text{yellow} \,  \sigma_z. 
		\eeq
		\ei 
		We remark that the choice of the names red, green, yellow and blue is just a convention, and, in this abstract setting, any name would be as valid as those used above. 
		Instead, what is really important to underline the role of $\sigma_z$ in this structure. $\sigma_z$ acts as a `\textit{chromatic antipodal operator}' in the sense that, given any pure hue state, its antipodal state is obtained by conjugation with $\sigma_z$ (given that $\sigma_z=\sigma_z^\dag=\sigma_z^{-1}$).
		
		\section{Conclusions and future perspectives}
		In this paper we have shown how to compute the perceptual attributes of perceived colors, initially defined in the model based on Jordan algebras and Hering's rebit, by means of the usual qubit density matrices and effects.
		%In this work we provided a proposal to bridge the gap between the real Jordan algebras framework of the quantum color perception model and the complex density matrix formalisms. 
		
		We hope that this will allow scientists familiar with the conventional algebraic language of quantum theories to have an easier access to that model which, analogously to quantum mechanics, underscores the pivotal role of the observer and measurement apparatus in comprehending color perception.  This motivates the need of this work, which is a step towards enabling an experimental verification of the theory.
		
		%Our structure presents a novel framework for human color perception, grounded in established principles while circumventing the constraints of metameric reduction. By employing Pauli matrices, we provide a geometric interpretation utilizing the Bloch sphere, thereby establishing a rigorous foundation for existing colorimetric concepts.

		In future works, we plan to perform a series of experiments which will necessarily involve chromatic opposition.  As an example, we can test entanglement effects associated to center-surround light stimuli in the following way:  from the center stimulus, a minimal number of photons are sent within a specific time interval to a cone area, activating the observer's action potentials;  simultaneously, other photons are sent from the surround. The physical properties of the center and its surround can be encoded in the entangled system of color perception.
		
		Specifically, we envision a density operator representing, say, the blue-yellow and yellow-blue Bell state, where the first subsystem corresponds to the center and the second to its surround. By measuring the composite system, we expect a blue center in a yellow surround 50\% of the time and a yellow center in a blue surround the other 50\%. If the surround stimulus is measured before the center stimulus arrives, then we expect an achromatic sensation for the center.
		
		While the results of such experiments could be explained by the quantum model, they pose a significant challenge to the classical theory of color perception.

\end{document}